\documentclass[twocolumn,superscriptaddress]{revtex4-2}
	\usepackage[colorlinks=true, a4paper=true, pdfstartview=FitV,
linkcolor=blue, citecolor=blue, urlcolor=blue]{hyperref}

\usepackage{amsmath}
\usepackage{amssymb}
\usepackage{graphicx}
\usepackage{hhline}
\usepackage{textcomp}
\makeatletter
\usepackage{babel}
\newcommand{\bea}{\begin{eqnarray}}
\newcommand{\eea}{\end{eqnarray}}

\newcommand{\be}{\begin{equation}}
\newcommand{\ee}{\end{equation}}

\usepackage[active]{srcltx}
\begin{document}


\title{Sphaleron without shape mode and its oscillon}

\author{K. Oles}
\affiliation{Institute of Theoretical Physics, Jagiellonian University,
Lojasiewicza 11, Krak\'{o}w, Poland}

 \author{J. Queiruga}
\affiliation{Department of Applied Mathematics, University of Salamanca, Casas del Parque 2 and \\
Institute of Fundamental Physics and Mathematics,
University of Salamanca, Plaza de la Merced 1, 37008 - Salamanca, Spain}

\author{T. Romanczukiewicz}
\affiliation{Institute of Theoretical Physics, Jagiellonian University,
Lojasiewicza 11, Krak\'{o}w, Poland}

\author{A. Wereszczynski}
\affiliation{Institute of Theoretical Physics, Jagiellonian University,
Lojasiewicza 11, Krak\'{o}w, Poland}

\begin{abstract}
We find that an oscillon can possess a characteristic double oscillation structure even though it results in a decay of a sphaleron which does not have any positive energy vibrational mode. We show that dynamics of such an oscillon can still be captured by collective coordinates provided by the sphaleron. Namely, its unstable mode and its scaling deformation i.e., Derrick mode.
\end{abstract}
\maketitle

\section{Motivation}
It has recently been postulated that there is an intimate relation between two rather distinct, non-perturbative objects, i.e., between {\it oscillon} and {\it sphaleron} \cite{MR}. Both are spatially localized, non-linear excitations of a fundamental field, which in the simplest version is just a real scalar in (1+1) dimensions. Concretely, sphaleron is an unstable, saddle point solution of static equations of motion (EoM), which may or may not carry a non-zero value of a pertinent topological charge \cite{M}. Oscillon \cite{G}, is a surprisingly long-lived oscillatory time-dependent solution \cite{BM, HS, CGM}, whose stability does not rely on any topological argument \cite{GS}. 

These solutions exist in various physical contexts from fundamental electroweak theory, where the sphaleron is a crucial ingredient in an explanation of the baryon asymmetry in the nature \cite{KM}, to cosmic oscillons \cite{G-cosm, Amin, LT}. 

The sphaleron-oscillon relation unifies a sphaleron and its oscillon, i.e., an oscillon which arises in decay of the sphaleron. In this case, the evolution of the oscillon may be well explained by the dynamics of collective coordinates provided {\it entirely} by the sphaleron. These are amplitudes of the negative energy (unstable) mode and the positive energy (massive vibrational) mode of the sphaleron. A resulting collective coordinate model (CCM) gives a much better approximation than the standard Fodor expansion, especially for large amplitude oscillons, where a characteristic structure of double oscillations shows up. 

The natural question that arises is what happens with the postulated sphaleron-oscillon relation if the {\it sphaleron does not support any positive energy mode} in the linear perturbation problem. TThe aim of the present work is to investigate this issue. For that purpose we will consider a very simple version of the $\phi^4$ theory, which we call the reverse $\phi^4$ model. This theory supports an exact sphaleron which does not carry any shape mode. 

\section{The reverse $\phi^4$ model}
\begin{figure}
\includegraphics[width=1.0\columnwidth]{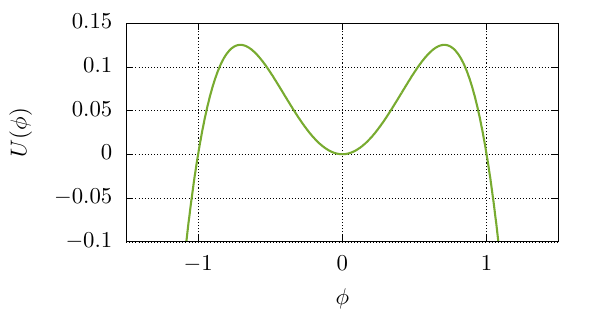}
 \caption{The reverse $\phi^4$ potential $U(\phi)$.} \label{potential-plot}
 \end{figure}

In this work we will focus on a real scalar field theory in (1+1) dimensions 
\be
L=\int_{-\infty}^{\infty} \left( \frac{1}{2}\phi_t^2 -\frac{1}{2}\phi_x^2 - V(\phi) \right)dx,
\ee
where the potential is a version of the $\phi^4$ model, which for obvious reasons we will call {\it the reverse  $\phi^4$ potential}
\be
V(\phi)=\frac{1}{2} \phi^2 (1-\phi^2), 
\ee
see Fig. \ref{potential-plot}.
This theory has one local minimum (false vacuum) with a perturbative excitation (pion) of the mass $m=1$. Furthermore, the potential is unbounded from below. Importantly, for $\phi \in [-1,1]$ the potential takes non-negative values with two additional zeros reached at $\phi=\pm 1$. The equation of motion reads
\be
\phi_{tt}-\phi_{xx}+\phi-2\phi^3=0.
\ee

This theory possesses two sphalerons
\be
\phi_S(x)= \frac{1}{\cosh (x-a)}, \;\;\; \phi_{\bar{S}}(x)= - \frac{1}{\cosh (x-a)},
\ee
where $a\in \mathbb{R}$ is their position. They differ by a multiplicative sign and therefore can be treated as a sphaleron and antisphaleron. As usual, the sphaleron piece-wisely solves the pertinent Bogomolny equations \cite{semi-BPS-1, semi-BPS-2}
\be
\frac{d\phi}{dx} = \pm \sqrt{2V} = \pm \phi \sqrt{1-\phi^2}.
\ee
Note that substitution $\psi = \phi^2$ brings this equation to the Bogomolny equation for the $\phi^3$-type model and therefore, our sphaleron is, up to a multiplicative constant, a square root of the cubic sphaleron \footnote{We thank N. Manton for brigning our attention to this fact. }.

The sphaleron solution has a very simple and analytically exact spectral structure. As usual the mode equation can be found by inserting of the perturbed static solution, $\phi(x,t)=\phi_S(x)+\eta(x,t)$, into the full EoM keeping only linear terms $\eta$. Next, we assume a periodic time-dependence of the perturbation, $\eta(x,t)=\eta(x)e^{i\omega t}$. Here we find (without loss of generality, we fix the position of the sphaleron at the origin)
\be
-\eta_{xx} + U(x)\eta = \omega^2 \eta,
\ee
where the Schrodinger-like potential 
\be
U(x)=V_{\phi\phi} (\phi_S) = 1-\frac{6}{\cosh^2x}
\ee
is the P\"oschl-Teller potential with $\lambda=2$. Hence, it supports one negative mode 
\be
\eta_{-1} \sim \frac{1}{\cosh^2 x} , \;\;\; \omega_{-1}^2= -3
\ee 
and one zero mode which reflects the translation symmetry of the theory
\be
\eta_0 \sim \frac{\sinh x}{\cosh^2 x} , \;\;\; \omega_0=0.
\ee
There are no other bound modes. In particular, the sphaleron does not have any positive energy discrete modes, the so-called shape modes. The mass threshold is at $\omega=1$ above which the continuum spectrum begins.

\section{The oscillon}
Let us now analyze properties of a large amplitude oscillon in the reverse $\phi^4$ theory. Obviously, the first question is whether it exits and, if this is the case, whether, despite of the nonexistence of shape modes of the sphaleron, it possesses a double quasi-periodic structure.

\begin{figure}
\includegraphics[width=1.0\columnwidth]{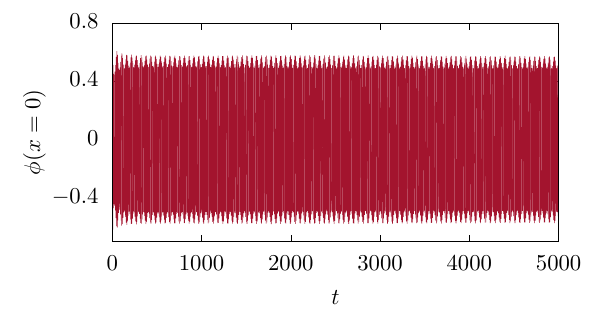}
\includegraphics[width=1.0\columnwidth]{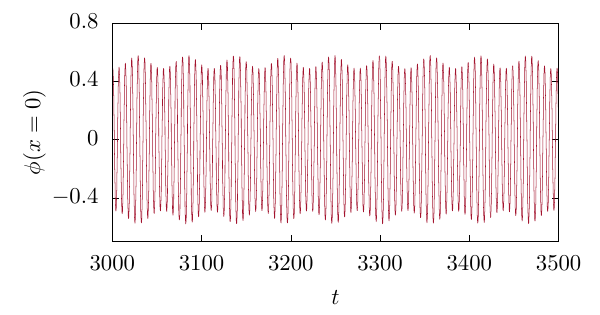}
\includegraphics[width=1.0\columnwidth]{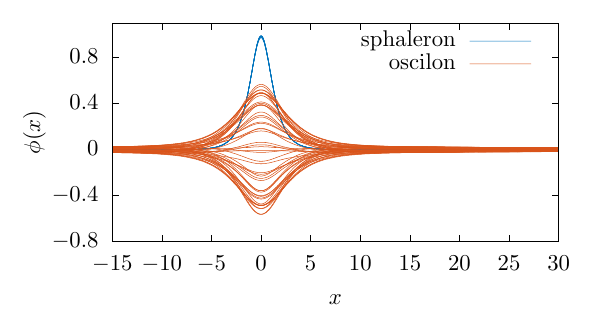}
 \caption{Oscillon. Upper/central: field at the origin $\phi(x=0,t)$. Lower: profiles of the field $\phi(x)$ at $t \in [3000,3500]$.} \label{oscillon-plot}
 \end{figure}

As an initial configuration we consider the sphaleron which is perturbed along the unstable direction such that it decays to the false vacuum. Specifically, 
\be 
\phi_{in}(x,0)=\frac{1}{\cosh(x)} + \frac{A}{\cosh^2(x)}, \label{init-sph}
\ee 
where the amplitude of the unstable mode is $A=0.01$. This initial configuration produces a {\it very stable} large oscillon which lives for a very long time with rather negligible radiative energy loss, see Fig. \ref{oscillon-plot}, upper panels. Importantly, we clearly see a {\it double quasi-periodic structure}. The fundamental frequency is $\omega_f=0.83$ while the frequency of the rumble is $\omega_r=0.126$. We underline that the existence of such a structure has recently been attributed to the existence of a shape mode of the underlying sphaleron, see the $\phi^3$ model \cite{MR}. However, in our case the sphaleron does not have such a mode and therefore this explanation cannot be valid any longer. 

We also remark that profiles of the oscillon are not too well bounded by the sphaleron, see Fig. \ref{oscillon-plot}, lower panel. Indeed, the oscillon extends significantly beyond the sphaleron, approaching the false vacuum in a slower manner. 
 
The observed double quasi-periodic structure of the evolution of large oscillons cannot be either described within the standard small amplitude expansion by Fodor et. el. \cite{Fod}. To see this, we perform the scaling transformation $\xi = \epsilon x$, $\tau=\omega t$, where $\omega=\sqrt{1-\epsilon^2}$. This brings the EoM to the following form
\be
(1-\epsilon^2) \ddot{\phi} - \epsilon^2 \phi''+\phi-2\phi^3=0. \label{Fodor_eom}
\ee
Now, as proposed in \cite{Fod}, we expand the field $\phi$ in a power series of $\epsilon$
\be
\phi=\epsilon \phi_1+\epsilon^2 \phi_2+\epsilon^3 \phi_3+...
\ee
Inserting it into (\ref{Fodor_eom}) we get for the first three terms
\bea
\phi_1&=&\frac{2}{\sqrt{3}} \frac{\cos \tau}{\cosh \xi}, \;\;\; \phi_2=0, \\
 \phi_3&=&\left( \frac{1}{18\sqrt{3}} \frac{\cosh 2\xi}{\cosh^3 \xi}   \right) \cos \tau - \frac{1}{6\sqrt{3}} \frac{\cos 3\tau}{\cosh^3 \xi}. \nonumber
\eea
One can easily verify that this approximation does not reproduce the double quasi-periodic behaviour of large oscillons. 
\section{Sphaleron based CCM model}
\begin{figure}
\includegraphics[width=1.0\columnwidth]{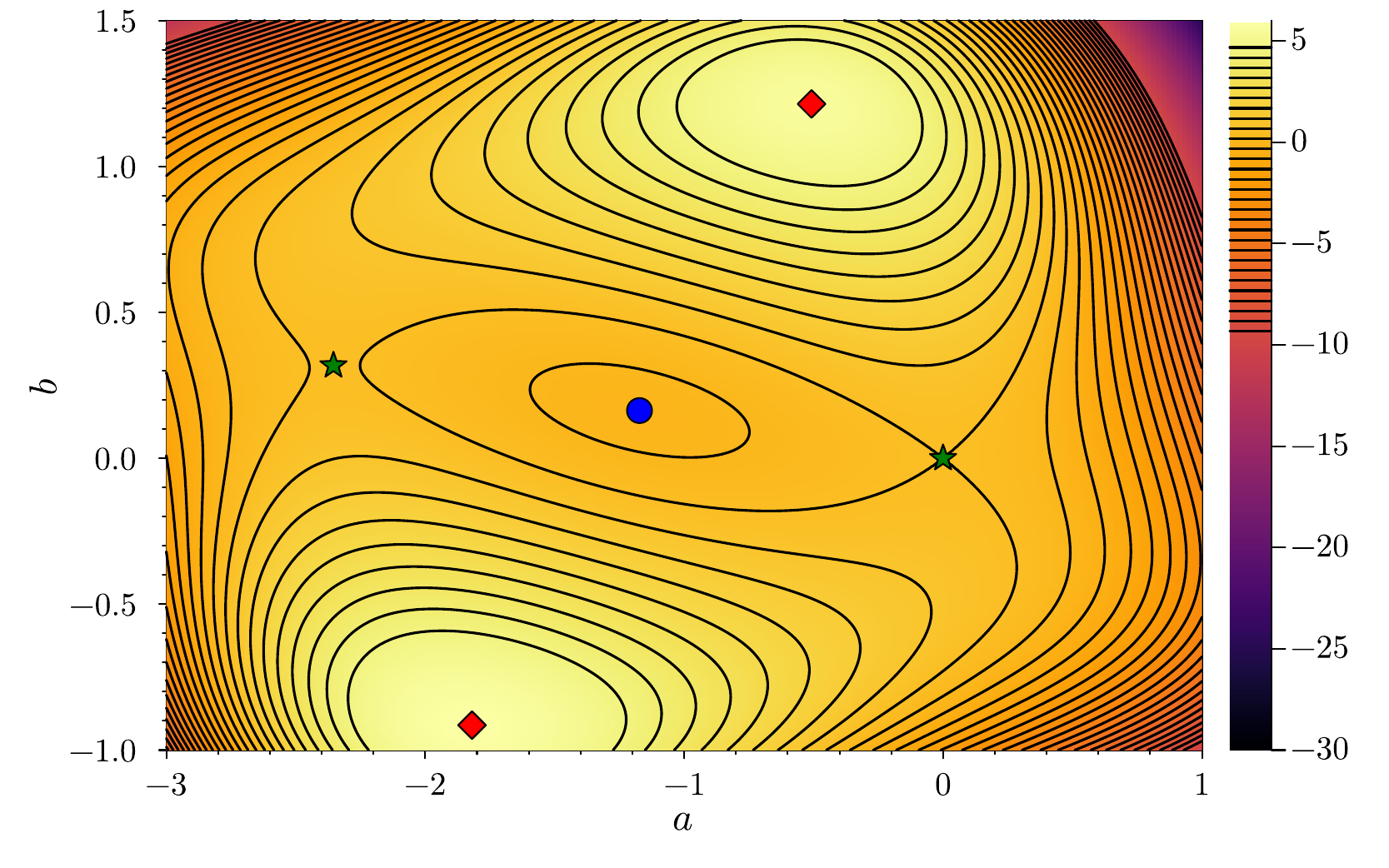}
 \caption{Effective potential $V_{eff}(a,b)$. The stars denote saddle point i.e, sphaleron configurations, the dot is the local minimum i.e., the false vacuum and the diamonds are local maxima. } \label{eff-potential-plot}
 \end{figure}

Since the sphaleron does not support any positive energy vibrational mode it is not clear whether the sphaleron-oscillon relation holds in this rather very simple and generic case. However, there is a kind of vibrational deformation which does exist for {\it any sphaleron} with or without shape modes. This is Derick mode which describes a scaling deformation $x \to \Lambda x$ with $\Lambda \in \mathbb{R}_+$. Since the unstable mode is obtained in the small perturbation procedure it is consistent to consider only a small scaling deformation $x \to (1+\epsilon)x$, where $|\epsilon| \ll 1$. Then,
\be
\phi_S(x) \to \phi_S((1+\epsilon) x) = \phi_S(x) - \epsilon \frac{x \sinh x}{\cosh^2 x} +o(\epsilon). \label{scale}
\ee
Hence the (first)Derrick mode is 
\be
\eta_D \sim \frac{x \sinh x}{\cosh^2 x}.
\ee
It has two zeros (if we count multiplicities) and therefore it is a natural next deformation to build a CCM.

We remark that this construction, called perturbative Relativistic Moduli Space (pRMS) approach, has recently proved to be very fruitful in the context of the collective coordinate description of kink-antikink collisions \cite{RMS}. In particular, it provides a converging scheme for the calculation of the critical velocity which divides quasi elastic, one-bounce scattering from non-elastic scenarios (e.g., annihilation or multi-bounces) \cite{RMS-2}.  

\begin{figure}
\includegraphics[width=1.0\columnwidth]{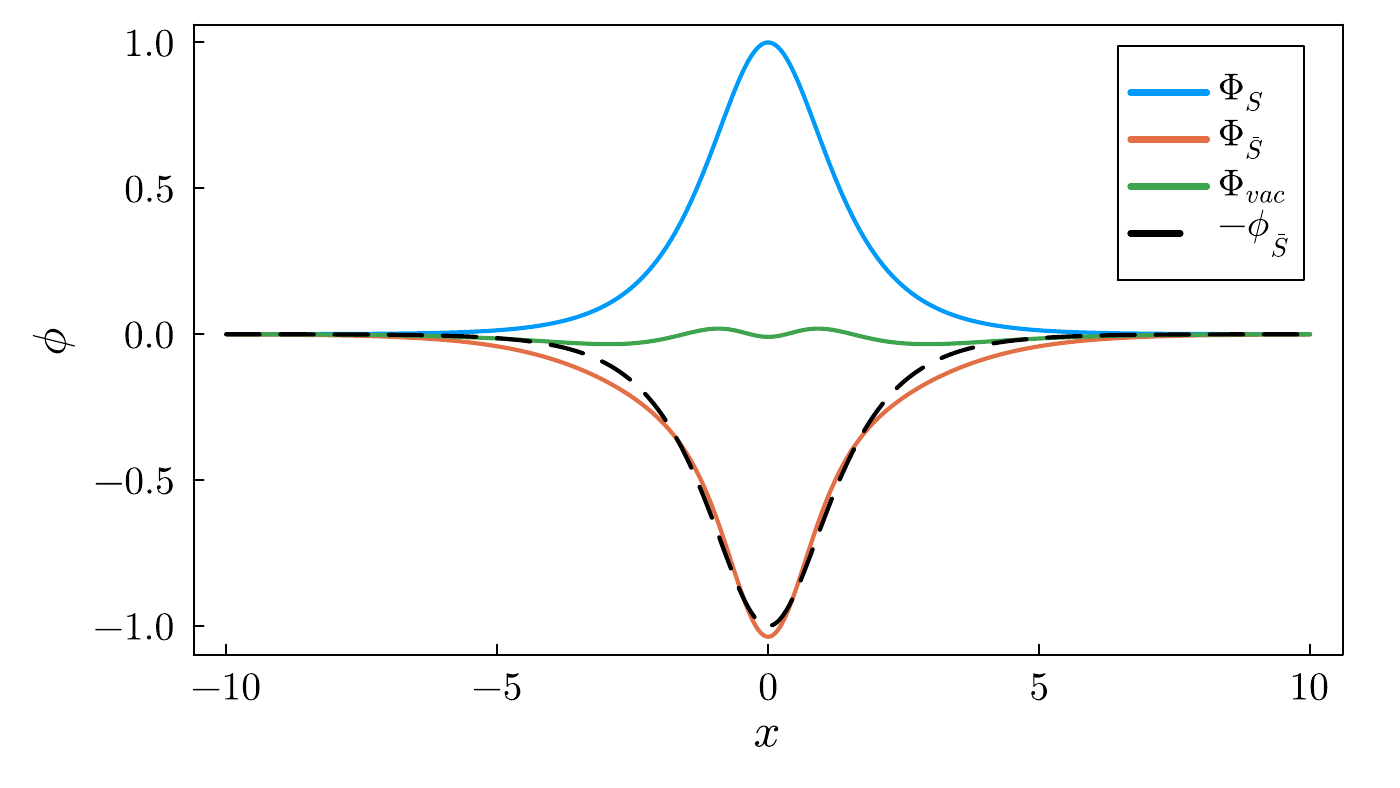}
 \caption{Profiles of the restricted set of configurations corresponding to the (anti)sphaleron and the false vacuum.} \label{plot-approx}
 \end{figure}
Based on this discussion we could propose the following restricted set of configurations  
\be
\Phi(x;a,b)= \frac{1}{\cosh x} +A \frac{1}{\cosh^2 x} + B\frac{ x \tanh x}{\cosh x},
\ee
where the amplitudes of the unstable mode of the sphaleron and its first Derrick mode serve as collective coordinates $A,B$. However, it is convenient to chose another, orthogonal basis of deformation. Using the Gram-Schmidt procedure we find 
\be
\Phi(x;a,b)= \frac{1}{\cosh x} +a \frac{1}{\cosh^2 x} + b \left( \frac{1}{\cosh^2 x}  - \frac{8}{\pi}  \frac{ x \tanh x}{\cosh x} \right). \label{1-der}
\ee
Now, we insert it into the original theory and promote the amplitudes $a,b$ to time-dependent variables. After integration over the spatial coordinate, we find a collective coordinate model 
\be
L[X^i] = \frac{1}{2} g_{ij} \dot{X}^i \dot{X}^j - V_{eff}(X^i),
\ee
where we identify the {\it moduli space} metric 
\be
g_{ij} (X^i) = \int_{-\infty}^\infty \frac{\partial \Phi}{\partial X^i} \frac{\partial \Phi}{\partial X^j} dx
\ee
and the effective potential 
\be
V_{eff}(X^i)=  \int_{-\infty}^\infty \left( \left( \frac{\partial \Phi}{\partial x}  \right)^2 + V(\Phi) \right) dx. 
\ee
Here, $(X^1,X^2)\equiv (a,b)$. In our case, the metric is diagonal 
\be
g_{aa}= \frac{4}{3}, \;\;\; g_{bb}=\frac{4}{9} \left( 5 + \frac{96}{\pi^2} \right), \;\;\; g_{ab}=0
\ee
and the potential reads 
 \begin{figure}
\includegraphics[width=1.0\columnwidth]{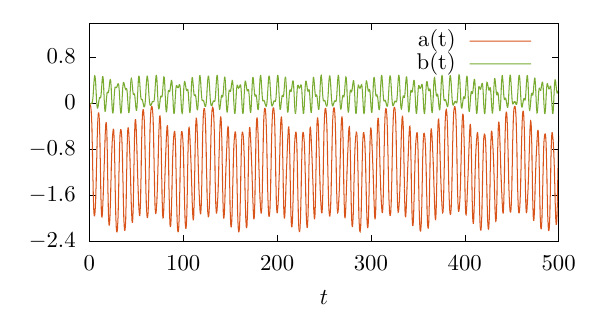}
\includegraphics[width=1.0\columnwidth]{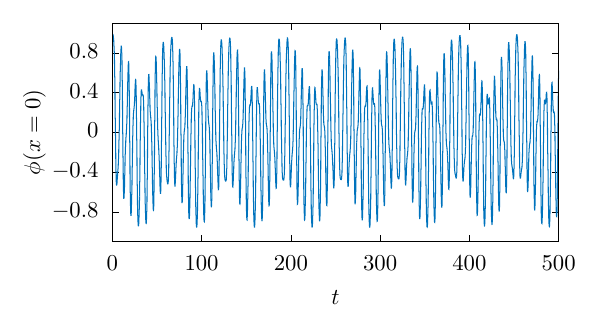}
\includegraphics[width=1.0\columnwidth]{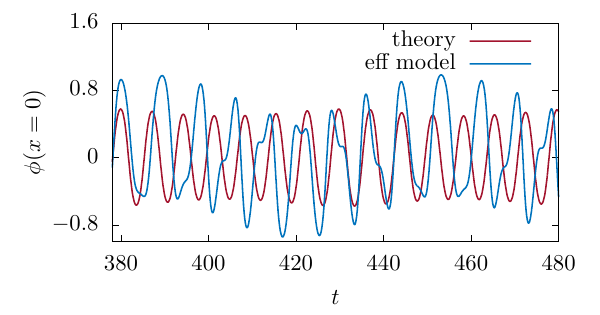}
 \caption{Dynamics in the CCM with the perturbed spheleron initial data (\ref{init-CCM-1}).} \label{CCM-plot}
 \end{figure}
  \begin{figure}
\includegraphics[width=1.0\columnwidth]{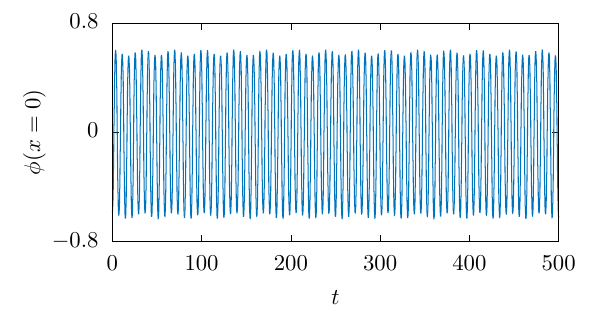}
\includegraphics[width=1.0\columnwidth]{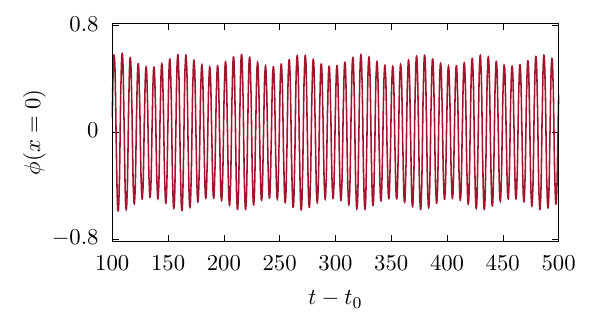}
\includegraphics[width=1.0\columnwidth]{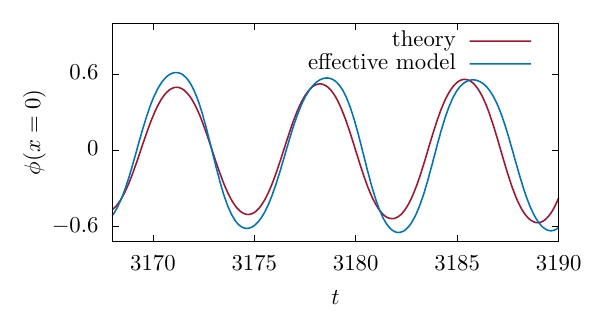}
 \caption{Comparison of dynamics in the CCM (upper) and full theory (medium) for the initial data corresponding to a quasi-stable oscillon (\ref{init-CCM-2}).} \label{CCM-plot-2}
 \end{figure}
\bea
V_{eff}(a,b)&=&
\frac{64 b^2 \left(6 a^2-126 a b-164 b^2+15\right)}{45 \pi ^2} \nonumber \\
&+& \frac{16 b \left(24
   a^2+171 a b+259 b^2\right)}{45 \pi } \nonumber \\
   &-&\frac{1}{8} \pi  (a+b) \left(5 a^2+10 a b+53
   b^2\right) \nonumber \\ &+& 
   \frac{1}{525} \left(-585 a^3 b+5 \left(210-319 a^2\right) \right. b^2 \nonumber \\
   &-&\left. 30 a^2
   \left(8 a^2+35\right)+9845 a b^3+7511 b^4+350\right) \nonumber \\
   &+& \frac{8192 b^4}{15 \pi
   ^4}-\frac{512 b^3}{3 \pi ^3},
   \eea
see Fig. \ref{eff-potential-plot}.
 
Of course, the effective potential has a saddle point at $a=b=0$ exactly reproducing the sphaleron. Here $V_{eff}(0,0)=2/3$. In addition, there is another saddle point at $a=-2.3544, b=0.3177$, which quite well approximates the antisphaleron. The effective potential at this point is $V_{eff}=V=0.686353$, which is a little higher than at the first saddle point. This is, in fact, a very nice and nontrivial feature that expansion around the sphaleron captures the antisphaleron. There is also a shallow local minimum at $a=-1.1724, b=0.1635$ which corresponds to the false vacuum, see Fig. \ref{plot-approx}. Here, $V=0.005913$. Importantly it supports bounded orbits which may model the oscillon.
 
Indeed, we evolved the CCM with initial conditions corresponding to the previous evolution of the full PDE 
   \be
   a(0)=0.01,  \;\; \dot{a}=0,\;\; b(0)=0,\;\; \dot{b}(0)=0. \label{init-CCM-1}
   \ee
The resulting oscillating trajectory is plotted in Fig. \ref{CCM-plot}. Specifically we show $a(t)$ and $b(t)$ as well as the value of the field at the origin, i.e., $\phi(x=0,t)=1+a(t)+b(t)$. It is clearly seen that the CCM reproduces the double frequency oscillations. However, the agreement is not perfect. Due to the flatness of the valley with the local minima the field oscillates between rather large values, reaching at the turning points $\pm 1$. This is significantly bigger than the extreme values found in the full PDE dynamics. Other important observables are the frequency of the fundamental oscillation $\omega_f$ and frequency of the rumbles $\omega_r$. In the CCM they read $\omega_f^{D} = 0.79$ and $\omega_r^{D}= 0.10$ while in the full theory we find $\omega_f=0.83$ and $\omega_r=0.13$. Here, the accordance is quite good.

The main reason for the observed disagreement is that the perturbed sphaleron in the ODE dynamics cannot radiate as there are no such dissipative DoF in the CCM model. The same initial state in the full theory settles down, via formation of an oscillon, which at the initial stage of the evolution radiates a significant amount of energy. Thus, in the CCM, this initial data rather provides a perturbed (wobbling) oscillon.

Because of that it is natural to compare the CCM with the PDE dynamics for initial conditions corresponding to a settled oscillon in the full theory. Specifically we chose a profile of the oscillon arising at late time evolution of the perturbed sphaleron (\ref{init-sph}) with $A=0.01$. Concretely, we take $t_0=3095.54$. At this time the oscillon reaches its returning point. Hence, in a good precision $\partial_t\phi(x,t_0)=0$. Now we find the best fit to such a profile in terms of our restricted set of configurations (\ref{1-der}). The least square fit gives
\be
a(t_0)=-1.9863, \;\;\; b(t_0)=0.3987. \label{init-CCM-2}
\ee
As the oscillon is chosen in its return point it is plausible to assume that the time derivatives vanishes $\dot{a}(t_0)=\dot{b}(t_0)=0$. In Fig. \ref{CCM-plot-2} we plot the results obtained in the CCM. It is clearly visible that now the oscillon is well modeled by the CCM evolution. For example, the modulation of the amplitudes of the field are smaller and agrees better with the PDE dynamics. The frequencies are now $\omega_f^{D}=0.85$ and $\omega^{D}_r=0.21$. The fundamental frequency is reproduced with very good precision while the frequency of the rumbles is too high. 

A similar improvement, when we change the initial data, has been noticed in the description of oscillon in $\phi^3$ theory \cite{MR}. 

All this verifies that the CCM based on the first Derick mode and the unstable mode quantitatively quite well describes the large amplitude oscillon. 

Three comments are in order. First of all, one may wonder what happens if we include  higher Derrick modes which enters via the expansion (\ref{scale}). In the case of kink-antikink collisions this improves some of CCM predictions as e.g., the critical velocity \cite{RMS-2}. In the case of the sphaleron the answer is not conclusive. In the simplest generalization we added the second Derrick mode which enters at $\epsilon^2$ order. Then we evolved the resulting three dimensional CCM  with the perturbed sphaleron initial data. This leads to a bound orbit, which, of course, is a welcome behaviour, however, with much more pronounced chaotic property. On the other hand if we start with initial conditions corresponding to a quasi-stable oscillon then, the CCM solution has rather tiny rumbles with too small modulations of the amplitude. We discuss this issue in appendix \ref{higher_derr}. 

Secondly, one may build an {\it at hoc} CCM with collective coordinates derived from the oscillon itself, see appendix \ref{osc-mod}. Concretely, they are an "amplitude" of the profile and the Derrick mode. An advantage of this set of configurations is that they extremely well reproduce the profiles of the true oscillon - also in the large $|x|$ regime. This wasn't the case for configurations based on the sphaleron DoF. Indeed, the tails of the sphaleron and oscillon differ quite a lot, see Fig. \ref{oscillon-plot}. Since this new CCM reproduces the oscillon amazingly well, it shows a direction in which our sphaleron based CCM should be modifed. Namely, one needs a deformation which allows for a better approximation of the tails. 

Thirdly, one may ask how good is the use of the first Derrick mode if the sphaleron possesses a shape mode. This happens e.g., in the $\phi^3$ model \cite{MR}. As we show it in appendix \ref{phi3_derr} the CCMs based on the first Derrick mode or on the shape mode are qualitatively equivalent. This originates from the fact that these modes are quite similar to each other. This feature was observed previously in the case of $\phi^4$ kink \cite{MORW}.
\section{Summary}
In the present paper we have studied the recently discovered sphaleron-oscillon relation in the case when the sphaleron does not have any positive energy bound mode. The existence of such a mode, together with the unstable mode of the sphaleron, was the main ingredient of the sphaleron-oscillon relation.  Indeed, it was shown that the dynamics of a large amplitude oscillon, which results in a decay of the sphaleron, can be quite accurately captured by a CCM containing entirely sphaleron anchored degrees of freedom. In particular, such a CCM describes double quasi-periodicity of the oscillon. 

The main result of this work is the extension of the validity of the sphaleron-oscillon relation in the cases where the sphaleron does not carry any positive energy bound mode. Indeed, rather surprisingly, we have found that in a model where the sphaleron has {\it no shape mode} the sphaleron-oscillon relation holds too. However, this requires a new, nontrivial ingredient. Namely, the {\it Derrick mode}. 

In the first step we have verified that large scale oscillons still enjoy double quasi-periodic structure. To explain its existence we have applied the first Derrick mode which describes a scale deformation of the sphaleron. Such a mode can be always defined for any sphaleron. The resulting CCM is again completely based on the sphaleron DoF and qualitatively captures the dynamics of large amplitude oscillon.

Furthermore, even in a model where the sphaleron has a shape mode it can be replaced by the first Derrick mode of the sphaleron in the CCM. This is not too surprising as both linear deformations are quite similar.  

It is intriguing that the existence of the shape mode of the sphaleron does not seem to be a crucial factor governing dynamics of the oscillon, e.g., the double quasi-periodic time dependence occurs regardless the sphaleron carries or not any shape mode. This is on the contrary to the kink collisions, where bound modes are the main actors in the famous resonant energy transfer phenomenon, which stands behind chaotic (or even fractal) structures observed in the final state formation \cite{Sug, CSW}. 

One possible explanation could be that sphalerons, also those without any bound modes, possess a hidden inner structure. Speaking more precisely, they might consist of constituents as for example a kink and antikink. Then motion of the constituents inside a sphaleron could provide internal DoF, captured e.g., by Derrick modes, which are further inherited by the corresponding oscillon, and which are responsible for its double oscillations. Undoubtedly, this idea should be further investigated. 

Interestingly, a good agreement between the sphaleron based CCM and the full field theory dynamics can be significantly improved if we correct the asymptotic of the profiles included into the collective coordinate construction. This has been done by construction of a CCM based on the oscillon itself. Of course, in the next step one should derive such a model from the underlying sphaleron.  

A CCM with sphaleron anchored collective coordinates provides also a natural path for the investigation of quantum oscillons. Due to the flatness of the moduli space it is rather straightforward to apply the canonical quantization of the underlying classical CCM. It would be interesting to compare results with other approaches \cite{H, V, ERW}. 
\section*{Acknowledgements}

K.O. acknowledges the support from the Polish National Science Centre 
(Grant NCN 2021/43/D/ST2/01122). 
J.Q. was supported by the Spanish Ministerio de Ciencia e Innovación (MCIN) with funding from the  European Union NextGenerationEU (PRTRC17.I1) and the Consejería de Educación, Junta de Castilla y León, through QCAYLE project, as well as MCIN Project No. PID2020-113406GB-I00 MTM. 

A.W. thanks Nick Manton for comments. 
\appendix
\section{Higher Derrick modes} \label{higher_derr}
\begin{figure}
\includegraphics[width=1.0\columnwidth]{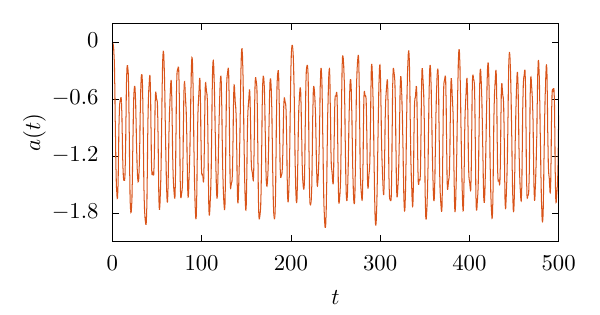}
\includegraphics[width=1.0\columnwidth]{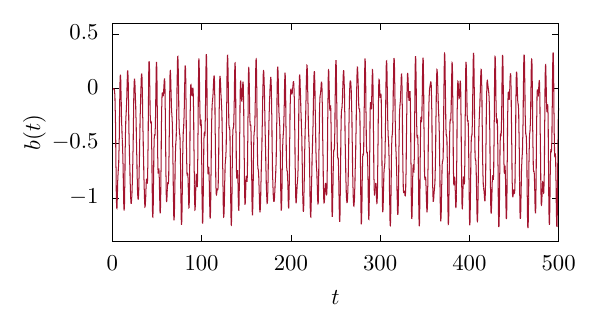}
\includegraphics[width=1.0\columnwidth]{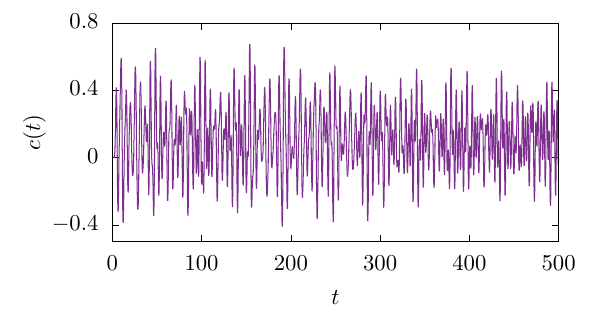}
\includegraphics[width=1.0\columnwidth]{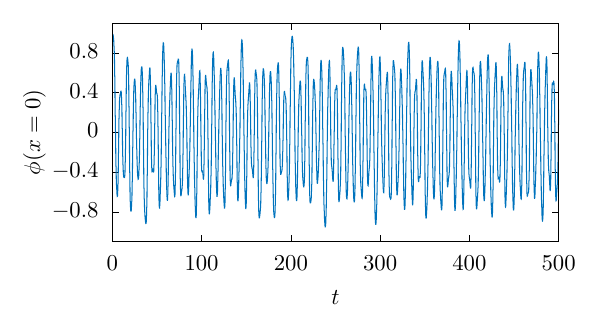}
 \caption{Dynamics in the CCM with the first two Derrick modes included and with the perturbed sphaleron initial data.} \label{2Der-plot}
  \end{figure}
  \begin{figure}
 \includegraphics[width=1.0\columnwidth]{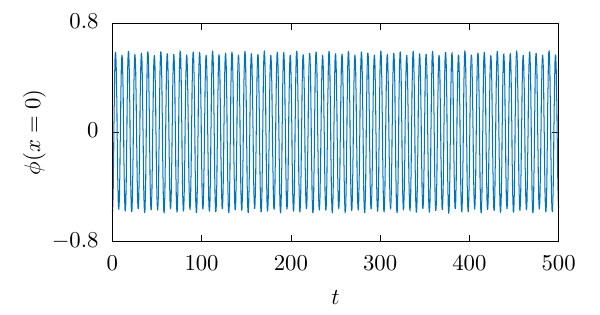}
  \caption{Dynamics in the CCM with the first two Derrick modes included and with the quasi-stable oscillon initial data (\ref{2-der-osc}).} \label{2Der-osc-plot}
 \end{figure}
The expansion (\ref{scale}) can be further continued. At the next order we find the second Derrick mode
\be
\eta_{2D} \sim  \frac{x^2}{\cosh x} \left(\tanh^2x -\frac{1}{2} \right).
\ee
It contributes to the following restricted set of configurations 
\bea
\Phi(x;a,b,c) &=& \frac{1}{\cosh x} + \frac{a}{\cosh^2 x}  \\ 
&+&  \frac{b x \tanh x}{\cosh x} + \frac{cx^2}{\cosh x} \left(\tanh^2x -\frac{1}{2} \right). \nonumber
\eea
For simplicity reasons we do not pass to the orthogonal basis. The resulting CCM has a constant but this time non-diagonal metric 
\be
g_{ij} = \left(
\begin{array}{ccc}
\frac{4}{3} & \frac{\pi}{6} & \frac{\pi}{3} - \frac{\pi^3}{32} \\
\frac{\pi}{6} & \frac{12+\pi^2}{18} & \frac{12+\pi^2}{24} \\
\frac{\pi}{3} - \frac{\pi^3}{32} & \frac{12+\pi^2}{24} & \frac{2}{5} + \frac{49 \pi^4}{7200} 
\end{array}
\right)
\ee 
and a rather complicated effective potential $V(a,b,c)$ which we do not show in explicit form. Here, $(X^1,X^2,X^3)\equiv (a,b,c)$.

We begin the analysis of dynamics of the corresponding CCM with the perturbed sphaleron initial data 
\be
a(0)=0.01, \;\dot{a}(0)=0, \;\;  b(0)=\dot{b}(0)=c(0)=\dot{c}(0)=0.
\ee
A welcome and still nontrivial feature is that the bounded orbit remains to be bounded. However, the obtained solution reveals much less regular character than in the CCM based on the first Derrick mode. For example, the clear double periodic structure seems to disappear and a much more chaotic dynamics is found. This is visible in Fig. \ref{2Der-plot}.

Next, we consider initial conditions corresponding to a quasi-stable oscillon. Again, we use the oscillon at $t_0=3095.54$ i.e., at a return point, where the kinetic energy is very small. Fitting the coordinates of the CCM to this profile we find
\be
a(t_0)=-1.588, \;\; b(t_0)=-1.013, \;\; c(t_0)=0.002. \label{2-der-osc}
\ee
Again we assume vanishing time derivatives. Now, the behavior of the field at the origin is completely different. We observe a very regular trajectory with fundamental frequency $\omega_f^{2D}=0.86$. However, we find too small modulation of the amplitude. 

In general, we saw that the dynamics of the CCM is extremely sensitive on the initial data. 

Although at the moment the results obtained in the CCM with two Derrick modes are not conclusive it seems that inclusion of next Derrick modes does not improve the effective description. This may be anchored in the fact that the oscillon has quite different asymptotic behavior with a significantly slower decay of the tails at large $|x|$. The oscillon is simply fatter than the sphaleron, see Fig. \ref{oscillon-plot}. Hence, the purely sphaleron based DoF cannot reproduce it unless quite many Derrick modes are incorporated.

\section{A CCM model from the oscillon and its first Derrick mode} \label{osc-mod}
Here we show an evidence which may help to explain why we do not see a clear improvement passing from the CCM based on one to two Derrick modes. Namely, this may be originated in the fact that this construction is based only on sphaleron DoF. Precisely specking, the sphaleron and its Derrick modes have rather different approach to the vacuum. Once it is corrected the resulting CCM leads to a much better agreement with the full PDE dynamics. This may be treated as a hint for future considerations. 

 \begin{figure}
 \includegraphics[width=1.0\columnwidth]{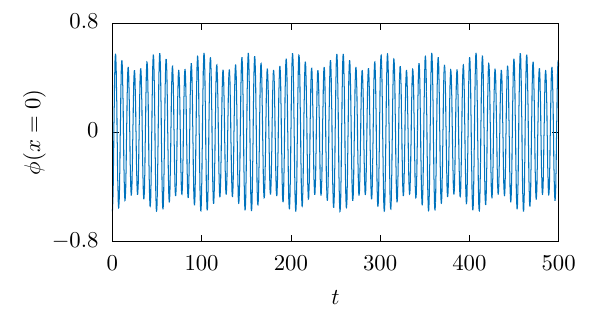}
  \caption{Dynamcis in the {\it ad hoc} CCM based the restricted set of configurations (\ref{ans-osci}.)} \label{osc-D-plot}
 \end{figure}
 
Here we propose a simple CCM based on degrees of freedom derived from the oscillon itself \cite{NNQ}. Therefore we approximate the profile of oscillon by the following function 
\be
\Phi = \frac{a_0}{\cosh \lambda_0 x },
\ee
where $a_0$ and $\lambda_0$ are fitting parameters. In the next step we treat them as dynamical coordinates $a,\lambda$. However, $\lambda$ can be once again related to a scaling deformation and therefore gives rise to Derrick modes via the usual Taylor expansion
\bea
\Phi &=& \frac{a}{\cosh (\lambda_0+\epsilon) x }   \\
&=& \frac{a}{\cosh(x\lambda_0)}+a \epsilon \frac{x \tanh(\lambda_0 x )}{\cosh(\lambda_0x)} +o(\epsilon).
\eea
Now, we promote each term in the expansion to a new independent deformation which amplitude defines an independent collective coordinate. Thus, at the second order in the expansion, we arrive at the following set of configurations 
\be
\Phi(x;a,b)= \frac{a}{\cosh(x\lambda_0)}+ b \frac{x \tanh(\lambda_0 x )}{\cosh(\lambda_0x)} , \label{ans-osci}
\ee
with $a$ and $b$ being collective coordinates. Inserting (\ref{ans-osci}) into the original Lagrangian and integrating over $x$ we get the following metric
\be
g_{ij}=\left(
\begin{matrix}
\frac{2}{\lambda_0} & \frac{1}{\lambda_0^2}\\
\frac{1}{\lambda_0^2} & \frac{12+\pi^2}{18\lambda_0^3}
\end{matrix}\right)
\ee
and the effective potential
\bea
V_{eff}(X^i)&=&-\frac{\left(7 \pi ^2-30\right) a b^3}{90 \lambda_0^4}-\frac{\pi ^2 a^2 b^2}{15 \lambda_0^3}-\frac{2 a^3 b}{3
   \lambda_0^2}\nonumber \\
  && +\frac{\left(3-\lambda_0^2\right)ab}{3\lambda_0^2}-\frac{2 a^4}{3 \lambda_0 }+\frac{\left(3+\lambda_0^2\right) a^2}{3 \lambda_0} \nonumber \\
   &&-\frac{\left(-240+20 \pi ^2+3 \pi ^4\right) b^4}{1800 \lambda_0^5}\nonumber \\
   &&+\frac{\left(7\pi^2\lambda_0^2+5(12+\pi^2)\right)b^2}{180 \lambda_0^3}
\eea
Here, $X^1=a$ and $X^2=b$.  Once again, we use the oscillon at $t_0=3095.54$ i.e., at a return point. Fitting the coordinates of the new CCM to this profile we find
\be
a(t_0)=-0.578, \;\; b(t_0)=0.096, \;\; \lambda_0=0.336.
\ee
This, together with the vanishing time derivatives define the initial data for the CCM. In Fig. \ref{osc-D-plot} we plot the value of the field at the origin resulting from this CCM. The agreement is spectacular. Furthermore, the frequencies also agree very well. Namely, we found $\omega_f^*=0.89$ and $\omega^*_r=0.15$, which are close to the true values. This shows that the correct profiles and their correct asymptotics are the crucial ingredients to build the correct CCM. It also proves the usefulness of the Relativistic Moduli Space framework. 
\section{The shape mode vs the first Derrick mode} \label{phi3_derr}
 
The first theory where the sphaleron-oscillon relation has been discovered was the $\phi^3$ model i.e., a scalar theory with a cubic potential
\be
V(\phi)=\frac{1}{2} \phi^2 - \frac{1}{3} \phi^3.
\ee
Again, the potential has a false vacuum at $\phi=0$ and an unbounded true vacuum at $\phi \to \infty$. The corresponding sphaleron
\be
\phi_S(x)=\frac{3}{2} \frac{1}{\cosh^2 \frac{x}{2}}
\ee
supports  an unstable mode 
\be
\eta_{-1}(x) \sim \frac{1}{\cosh^3 \frac{x}{2}}
\ee
and a shape mode
\be
\eta_{1}(x) \sim \frac{4\cosh^2 \frac{x}{2} -5 }{\cosh^3 \frac{x}{2}}.
\ee
These two deformations of the sphaleron led to a CCM based on the following restricted set of configurations \cite{MR}
\bea
\Phi(x;a,b)&=&\frac{3}{2} \frac{1}{\cosh^2 \frac{x}{2}} \\ 
&+&  a \sqrt{\frac{15}{32}} \frac{1}{\cosh^3 \frac{x}{2}} + b \sqrt{\frac{3}{32}}    \frac{4\cosh^2 \frac{x}{2} - 5}{\cosh^3 \frac{x}{2}}. \nonumber 
\eea
It was shown that this CCM captures the main aspects of the large oscillons in the full PDE dynamics \cite{MR}. 
\begin{figure}
\includegraphics[width=1.0\columnwidth]{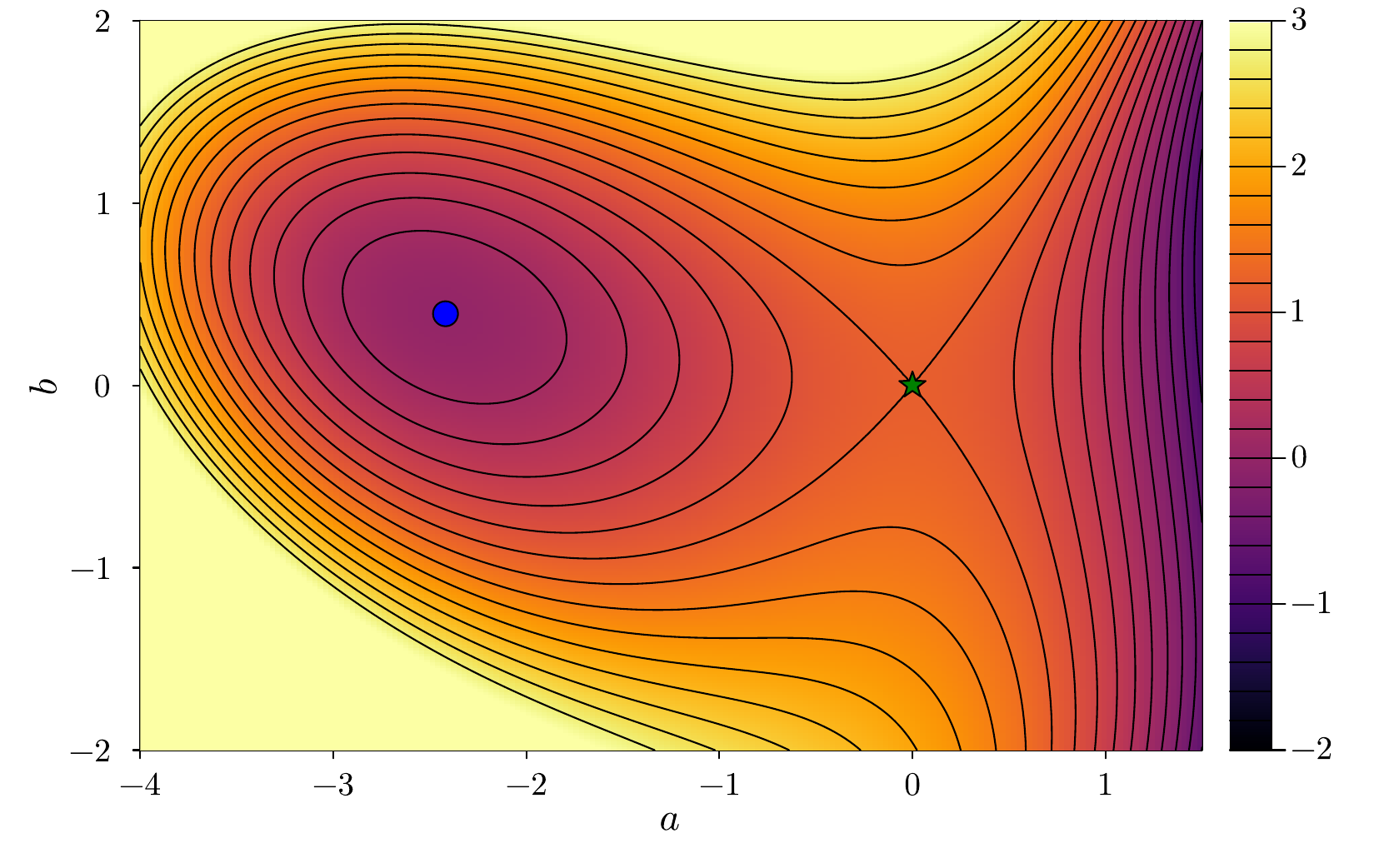}
 \caption{The effective potential $V_{eff}(a,b)$ for $\phi^3$ model with the first Derick mode.} \label{Veff-phi3-plot}
 \end{figure}

As it has been remarked one may replace the shape mode of the sphaleron by its first Derrick mode
\be
\eta_D (x) \sim \frac{x \tanh \frac{x}{2} }{\cosh^2 \frac{x}{2}}.
\ee
The pertinent restricted set of configurations (with orthogonal and normalized modes) reads
\bea
\Phi(x;a,b) &=& \frac{3}{2} \frac{1}{\cosh^2 \frac{x}{2}} + a \sqrt{\frac{15}{32}} \frac{1}{\cosh^3 \frac{x}{2}} \\
&+& b \sqrt{\frac{3645}{24992}} \left(  \frac{1}{\cosh^3 \frac{x}{2}} - \frac{64}{9\pi}  \frac{x \tanh \frac{x}{2} }{\cosh^2 \frac{x}{2}}\right). \nonumber
\eea
Now, the moduli space metric is just a unit matrix while the effective potential is
\bea
&&V_{eff}(a,b) = 
-\frac{1}{349776035840 \pi ^3} \times \\
&& \left( 35145 \sqrt{7810} \pi ^2 \left(11025 \pi ^2-65536\right) a^2 b \right. \nonumber \\
&& +4269727 \pi ^3
   \left(25 a^2 \left(35 \sqrt{30} \pi  a+2048\right)-98304\right)\nonumber \\
   && +  3905 \pi 
   \left(\sqrt{30} \pi  \left(7344225 \pi ^2-62652416\right) a \right.  \nonumber \\
   && -43008 \left. \left(4096+405 \pi
   ^2\right)\right) b^2 \nonumber \\
   &+&  5 \sqrt{7810} \left(8455716864-2890465280 \pi ^2 \right. \nonumber \\
   &&\left. \left.  +182219625 \pi
   ^4\right) b^3 \right). \nonumber\label{eff_pot_phi3_der}
   \eea
   
\begin{figure}
\includegraphics[width=1.0\columnwidth]{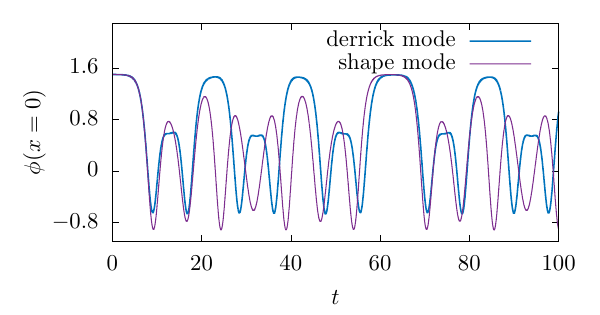}
 \caption{$\phi^3$ model. Comparison of dynamics obtained from the CCM based on the Derrick mode and the shape mode of the sphaleron.} \label{dyn-phi3-plot}
 \end{figure}
 
 \begin{figure} 
\includegraphics[width=1.0\columnwidth]{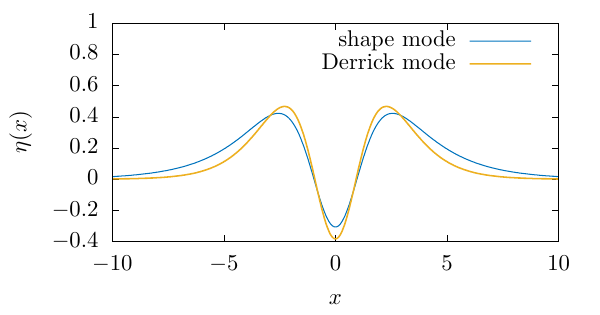}
 \caption{Comparison of the orthonormalized Derrick and the shape mode of the sphaleron in the $\phi^3$ model.} \label{modes-phi3}
 \end{figure}
 
We plot it in Fig. \ref{Veff-phi3-plot}. In Fig. \ref{dyn-phi3-plot} we present  dynamics obtained in the CCM based on the upper set of configurations. It reproduces the PDE dynamics qualitatively as good as the CCM based on the shape mode. This is because the effective potential (\ref{eff_pot_phi3_der}) looks very similar to the one obtained using the shape mode of the sphaleron \cite{MR}. This has its origin in the similarity between the shape mode and the orthonormalized first Derrick mode $\tilde{\eta}_D$. E.g., their overlap is close to 1
\be
\int_{-\infty}^\infty \eta_{1}(x) \tilde{\eta}_D(x) dx = 0.9701,
\ee
see also Fig. \ref{modes-phi3}, where both modes are plotted. Here,
\be
 \tilde{\eta}_D(x) = \sqrt{\frac{3645}{24992}} \left(\frac{64}{9\pi}  \frac{x \tanh \frac{x}{2} }{\cosh^2 \frac{x}{2}}-  \frac{1}{\cosh^3 \frac{x}{2}} \right). 
\ee


\begin{thebibliography}{99}

\bibitem{MR} N. Manton and T. Romanczukiewicz, The Simplest Oscillon and its Sphaleron, Phys. Rev. D 107 (2023) 085012.

\bibitem{M}  N. Manton, The Inevitability of Sphalerons in Field Theory,  Phil. Trans. Roy. Soc. Lond. A377 (2019) 2161, 20180327.

\bibitem{G} M. Gleiser, Pseudostable bubbles, Phys. Rev. D49 (1994) 2978.

\bibitem{BM} I. L. Bogolyubsky and V. G. Makhankov, On the pulsed soliton lifetime in two classical relativistic theory models, JETP Lett. 24 (1976) 12.

\bibitem{HS} M. Hindmarsh and P. Salmi, Numerical investigations of oscillons in 2 dimensions, Phys.Rev.D 74 (2006) 105005. 

\bibitem{CGM} E. J. Copeland, M. Gleiser, and H. R. Muller, Oscillons: Resonant configurations during bubble collapse, Phys. Rev. D52 (1995) 1920.

\bibitem{GS} M. Gleiser and D. Sicilia, A General Theory of Oscillon Dynamics, Phys. Rev. D 80, 125037 (2009)

\bibitem{KM} F. R. Klinkhamer and N. S. Manton, A saddle point solution in the Weinberg-Salam theory, Phys. Rev. D 30, 2212 (1984).

\bibitem{G-cosm} M. Gleiser, Oscillons in scalar field theories: Applications in higher dimensions and inflation, Int. J. Mod. Phys. D 16 (2007) 219.

\bibitem{Amin} M. A. Amin, R. Easther, H. Finkel, R. Flauger and M. P. Hertzberg, Oscillons After Inflation, Phys. Rev. Lett. 108, 241302 (2012).

\bibitem{LT} K. D. Lozanov, V. Takhistov, Enhanced Gravitational Waves from Inflaton Oscillons, Phys. Rev. Lett. 130 (2023) 18, 181002.

\bibitem{semi-BPS-1} A. Alonso-Izquierdo, S. Navarro-Obregon, K. Oles, J. Queiruga, T. Romanczukiewicz, and A. Wereszczynski, Semi-BPS sphaleron and its dynamics, arXiv:2308.14420. 


\bibitem{semi-BPS-2} N. S. Manton, Integration Theory for Kinks and Sphalerons in One Dimension, arXiv:2308.14453. 

\bibitem{Fod} G. Fodor, P. Forgacs, Z. Horvath, and A. Lukacs, Small amplitude quasibreathers and oscillons, Phys. Rev. D 78, 025003 (2008),


\bibitem{RMS} C. Adam, N. S. Manton, K. Oles, T. Romanczukiewicz, and A. Wereszczynski, Relativistic Moduli Space for Kink Collisions, Phys. Rev. D105 (2022) 065012.

\bibitem{RMS-2} C. Adam, D. Ciurla, K. Oles, T. Romanczukiewicz, and A. Wereszczynski, Relativistic Moduli Space and critical velocity in kink collisions, Phys. Rev. E108  (2023) 024221. 

\bibitem{MORW} N. S. Manton, K. Oles, T. Romanczukiewicz, and A.
Wereszczynski, Collective coordinate model of kink-antikink collisions in $\phi^4$ theory, Phys. Rev. Lett. 127 (2021) 071601.

\bibitem{Sug} T. Sugiyama, Kink-antikink collisions in the two-dimensional $\phi^4$ model, Prog. Theor. Phys. 61, 1550 (1979).

\bibitem{CSW} D. K. Campbell, J. F. Schonfeld, and C. A. Wingate, Res- onance structure in kink-antikink interactions in $\phi^4$ theory, Physica (Amsterdam) 9D, 1 (1983).

\bibitem{H} M. P. Hertzberg, Quantum Radiation of Oscillons, Phys. Rev. D 82 (2010) 045022.

\bibitem{V} J. Olle, O. Pujolas, T. Vachaspati and G. Zahariade, Quantum Evaporation of Classical Breathers, Phys. Rev. D 100 (2019) 045011. 

\bibitem{ERW} J. Evslin, T. Romanczukiewicz, A. Wereszczynski, Quantum Oscillons May be Long-Lived, arXiv:2305.18056, to appear in JHEP. 

\bibitem{NNQ} S. Navarro-Obregón, L.M. Nieto, J.M. Queiruga, Inclusion of radiation in the CCM approach of the $\phi^4$ model, arxiv:2305.00497. 


\end{thebibliography}
\end{document}